\documentclass[preprint]{elsarticle}

\usepackage{tabulary,graphicx,amsfonts,amsmath,amssymb,amsbsy,dcolumn,bm}
\usepackage[utf8]{inputenc}
\usepackage{lineno,hyperref}
\usepackage{chemformula} 
\usepackage[T1]{fontenc} 
\usepackage{graphicx}
\usepackage{bm}
\hypersetup{colorlinks = true, urlcolor = blue, linkcolor = blue, citecolor = blue}
\usepackage{color}
\usepackage[normalem]{ulem}
\usepackage{float}
\usepackage{subfigure}
\usepackage{amsmath}
\usepackage{graphicx}
\usepackage{amsfonts}
\usepackage{mathrsfs,relsize,array}
\usepackage{appendix}
\usepackage{romannum}
\usepackage{makecell}
\usepackage{multirow}
\usepackage[flushleft]{threeparttable}
\usepackage[mediumspace,mediumqspace,Grey,squaren]{SIunits}

\newcommand{\vg}[1]{\textcolor{black}{#1}}
\newcommand{\AT}[1]{\textcolor{black}{#1}}
\newcommand{\HL}[1]{\textcolor{black}{#1}}

\modulolinenumbers[5]

\journal{JOURNAL OF NON-CRYSTALLINE SOLIDS}

\begin{document}

\begin{frontmatter}
\title{A continuum model reproducing the multiple frequency crossovers in
acoustic attenuation in glasses}
\author[lamcos]{H. Luo}

\author[ilm]{V. M. Giordano}
\author[lamcos]{A. Gravouil}
\author[lamcos,onera]{A. Tanguy}
\address[lamcos]{Univ Lyon, INSA-Lyon, CNRS UMR5259, LaMCoS, F-69621, France}
\address[onera]{ONERA, University Paris-Saclay, Chemin de la Hunière, BP 80100, 92123
Palaiseau, France}
\address[ilm]{Institut Lumi\`ere  Mati\`ere, UMR 5306 Universit\'e Lyon 1-CNRS,  F-69622 Villeurbanne Cedex, France}

\begin{abstract}
Structured metamaterials are at the core of  extensive research, promising for acoustic  and thermal engineering. Nevertheless,  the computational cost  required
for correctly simulating large systems imposes to use a continuous model to describe the effective behavior without knowing the atomistic details. 
Crucially, a correct description needs to describe both the  extrinsic interface-induced and the intrinsic atomic scale-originated phonon scattering, especially when the component material   is made of glass, a highly dissipative material in which wave attenuation is strongly dependent on frequency as well as on temperature.
In amorphous systems, the effective acoustic attenuation triggered by multiple mechanisms  is now well characterized  and exhibits a  nontrivial frequency dependence with a double crossover of power laws.
In this work, we propose a  continuum viscoelastic model based on the hierarchical strategy multi-scale approach, able to   reproduce well the phonon attenuation in a large frequency range, spanning three orders of magnitude from GHz to THz
with a $\omega^2-\omega^4-\omega^2$ dependence, including the influence of  temperature. 
\end{abstract}


\end{frontmatter}


\section{Introduction}

Heterogeneous architectured materials have attracted considerable attention on their potential applications~\cite{Bouaziz2008}. They are manmade structural materials  designed for getting \textit{ad hoc} properties, which are not normally found in nature. 
Generally, these materials are achieved by engineering the interplay between the shape, the properties and spatial distribution of materials at different scales. 
The length scale of architectures ranges from nanometer to macroscopic sizes, depending on the application.  For example, \vg{acoustic} metamaterials are composite materials with  repeating regular patterns \vg{on a macroscopic scale}, which have been initially investigated for applications in acoustics, for realizing  acoustic guides, filters, lenses~\cite{Zhang2009,Cummer2016}. \vg{Engineered at a smaller scale, they} have also  been  introduced into the field of thermal science to  realize thermal cloaks and camouflage ~\cite{Sklan2018}, tunable multi-functional thermal metamaterials ~\cite{Park2017},  etc. 
In both applications, the goal is to filter, guide and prevent the propagation of sound waves (phonons), responsible  for the acoustic transport \AT{(in the sub-GHz range), and for the heat transport at room temperature (in the THz range). }
Heterogeneous architectured materials affect the propagation of the phonons mainly by introducing an extrinsic  scattering source, such as the interfaces.\AT{ In general, sound propagation in such heterogeneous structures displays non trivial effects.} For example, in the case of metamaterials featuring periodicity,  novel coherent effects arise due to the Bragg scattering. As a result, a new defined Brillouin Zone will  fold  the band structure, thus introducing new optic modes which can scatter the  acoustic ones. Furthermore,  the  elemental bricks in the micro-structure introduce local resonances, resulting in the presence of band gaps in the dispersion relation, inside which phonons cannot exist~\cite{Liu2000}, \vg{thus preventing sound or heat transport.}

Recently, it has been proposed to use amorphous materials as building blocks for metamaterials for thermal transport engineering, due to their renowned low thermal conductivity~\cite{France-Lanord2014,Zhu2018,Tlili2019,Hussein2019}. In particular, nanocomposites made of amorphous and crystalline components have shown to be promising for dramatically affecting thermal transport  \cite{Nakamura2018}, while preserving mechanical stability. 
To correctly describe the phonon propagation in complex structured  materials, it is needed to properly investigate the effect of heterogeneities and interfaces \vg{on a macroscopic scale. For this, simulations on very large systems are necessary, which are beyond the possibilities of atomistic simulations, but can be acheved through continuous modeling tractable in Finite Element calculations. This requires  to model the detailed microstructure on large systems, and, at the same time, to be able to represent accurately the main characteristics of phonons propagation in the GHz to THz frequency range. For that, it is fundamental to take properly into account the intrinsic phonon scattering sources of the composing materials at play in the same frequency range. This is even more important when one of these materials} is an amorphous phase, a highly dissipative material, where phonon attenuation changes with temperature and phonon frequency, following different power laws depending on the frequency range. 
In order to provide a realistic description of the effect of the nanostructure on phonon propagation, a multiscale approach needs thus to be adopted, \vg{where both intrinsic attenuation with atomistic origin and extrinsic attenuation induced by the interfaces are present. In this article, we will propose a continuous model able to describe intrinsic attenuation in glasses. The model we propose involves three parameters that can be adjusted depending on the chemical composition of the glass, and on the temperature.}

In glasses, apparent acoustic attenuation  can be related to a viscous consumption of mechanical energy induced by anharmonic effects~\cite{Schirmacher2010}, or it can result from acoustic scattering in disordered harmonic systems~\cite{Beltukov2016,Gelin2016,Damart2017,Beltukov2018}. It can be temperature-dependent, or not\cite{Ruocco1999,Baldi2010,Ruffle2003,Ruffle2006,Monaco2009, Mizuno2020}. Most interestingly, \AT{unlike in usual continuous models of sound attenuation, acoustic attenuation in glasses is frequency-dependent\cite{Baldi2010,Ayrinhac2011}.}
Numerous experiments and numerical works (atomistic simulations) have investigated  the acoustic attenuation in the GHz to THz frequency range \cite{Baldi2010,Ruffle2003,Ruffle2006,Monaco2009,Vacher1997,Monaco2009a,Ruocco1999,
Masciovecchio2004,Benassi2005,Masciovecchio2006,Devos2008,Dietsche1979,Buchenau2014,Mizuno2020,Wang2019,Gelin2016,Scopigno2003,Fioretto1999,Benassi1996}. 
A non-exhaustive summary of experimental and theoretical literature results on acoustic attenuation in glasses is reported in Tab.~\ref{tab:sound_attenuation_literatre_1} and Supplementary Material Tab. S1.
Numerous attenuation channels are involved at the atomic scale, the importance of which depends on the phonon frequency and temperature: anharmonicity of the interatomic interactions~\cite{Schirmacher2010}, thermally activated relaxation processes~\cite{Ayrinhac2011}, tunneling due to two level {systems~\cite{Gilroy1981,Phillips1987,Parshin2007}, soft modes~\cite{Buchenau1992,Ji2019}, and scattering resulting from structural disorder~\cite{Beltukov2016,Gelin2016,Beltukov2018}. Especially, this last contribution is independent on temperature and generally dominates phonon attenuation in the GHz to THz frequency range~\cite{Damart2017}.}
In the low frequency range, anharmonicity leads to a phonon attenuation, $\Gamma$, which scales as $\omega^a$ with $a\approx 1.5-2$ as shown in molecular dynamics simulations~\cite{Mizuno2020}. Near the THz range, \textit{i.e.}  for nanometric phonon wavelengths, scattering changes progressively from weak to strong and  scattering due to the atomistic disorder starts to dominate, exhibiting a Rayleigh-like dependence $\Gamma \propto \omega^4$ that can be computed analytically~\cite{Gelin2016,Tanguy2010}. Finally, there is again a new high frequency $\Gamma \propto \omega^2$ regime~\cite{Ruffle2003,Ruffle2006,Ruffle2008,Levelut2006,Monaco2009,Baldi2010,Duval1998,Ayrinhac2011,Mizuno2014}, eventually followed by a sudden drop \cite{Damart2017}. The transition to strong scattering takes place at frequencies near the Boson Peak and has been interpreted as resulting from nanometric elastic heterogeneities \cite{Tanguy2002,Leonforte2006,Tanguy2010,Duval1998,Mizuno2014,Marruzzo2013,Schirmacher2015b,Conyuh2020}. 
At low temperature, MD simulations are capable to reproduce these attenuation crossovers, with various kinds of scaling rules~{\cite{Monaco2009a,Marruzzo2013,Beltukov2016,Gelin2016,Wang2019}}. At finite temperature, the consideration of anharmonicity  has finally drawn a complete picture of phonon attenuation in glasses ~{\cite{Ruffle2008,Mizuno2020}}. \AT{We can thus finally classify three regimes of acoustic attenuation vs frequency in glasses:} (1) $\Gamma \propto \omega^{a}$ due to the anharmonicity at low frequencies, with $a \approx 1.5-2$ and temperature-dependent; (2) $\Gamma \propto \omega^4$ due to {Rayleigh}-like scattering induced by structural disorder. During this regime the collective vibrational modes lose progressively their plane wave character; (3) $\Gamma \propto \omega^2$ above Ioffe-Regel \vg{crossover} ($\omega_{IR}=\pi \Gamma$), where phonons can no longer be considered as propagative plane waves. After that,  they propagate in a diffusive way, and even may localize at higher frequency(in the Anderson's definition of localization, but for acoustic waves)~\cite{Allen1999,Beltukov2016,Beltukov2018,Skipetrov2018,Page2009}. 

Nevertheless, little consideration has been given so far to the use of those results  on acoustic attenuation at the macroscale due to the lack of adapted continuum constitutive laws.  
As said, a continuous
modeling of phonon attenuation in an amorphous-based nanocomposite should also describe the frequency and temperature dependence of
attenuation in glasses. This requires to bridge  atomistic and macroscopic scale, while keeping information on intrinsic attenuation channels active as a function of frequency and temperature. 
To do so,  a model is urgently needed to homogenize the frequency-dependent effective attenuation triggered by various mechanisms.
Recently, we have proved  the  concept of using a continuum viscoelastic model, characterized by the quality factor $Q^{-1} = G''/G'$ with $G= G' + iG''$  the complex elastic modulus in the linear regime~\cite{Gilroy1981}, to describe sound damping in an amorphous material at THz frequencies up to the Ioffe-Regel \vg{crossover}\cite{Luo2020}. Within the Maxwell theory, such a viscosity is related to the stress  relaxation process, giving a lower boundary of viscosity for structural relaxation kinetics \cite{Lancelotti2021}.
Considering the exponential decay of acoustic waves in both viscoelastic medium and glass up to Ioffe-Regel frequency, a relation can be derived between the microscopic ($\Gamma$) and  the macroscopic quantity ($Q^{-1}$) ~\cite{Pohl2002,Damart2017,Schirmacher2013,Schirmacher2015b,Parke1966,Carfagni1998} : 
\begin{equation}\label{eq:relation}
  \Gamma/\omega = Q^{-1}  
\end{equation}
able to interpret experimental neutron-scattering data for example, within the Damped Harmonic Oscillator Model for phonons \cite{Chen2017a}.
In that work,  we have assumed a constant viscosity, thus leading to a single $\omega^2$ behaviour in the acoustic attenuation. However, in such frequency range,  the atomistic structural disorder dominates the scattering \cite{Mizuno2014,Damart2017} thus leading to a $\omega^4$ behaviour, as found in ~\cite{Ruffle2006,Monaco2009,Baldi2010,Ayrinhac2011,Mizuno2020}.
The corresponding scattering  is potentially responsible for the unusual temperature dependence $\kappa\propto T$ observed in the low temperature regime \cite{Tlili2019}, while this low temperature sensitivity was also related to anharmonic effects in the 2-level model 
\cite{Gilroy1981,Phillips1987,Buchenau1992}. The progressive transition to a stronger scattering and to the resulting diffusive motion of initial plane waves will be responsible for the progressive saturation which may be observed as the peak in the specific heat   ~\cite{Zeller1971,Pohl2001,Tlili2019} and a plateau in the glassy thermal conductivity at around 10 K followed by an increase of $\kappa$ with the temperature. 
In order to model amorphous materials for thermal applications, it  is crucial to include different power laws of frequency,  and more specifically, at least  three successive regimes including $\Gamma\propto\omega^a$ with $a\approx 2$ to take account of \AT{various} anharmonic effects in the low frequency range at different temperatures \cite{Mizuno2020}) , $\Gamma\propto\omega^4$ in the low scattering regime and then again, at very high frequencies,  $\Gamma\propto\omega^2$  in the strong scattering (diffusive) regime.

In this work, we develop  a  novel continuum viscoelastic model, without introducing disorder, by taking into account the different attenuation regimes acting in parallel in the rheological responses of glasses. \AT{We call it the Parallel Power Law Model (PPLM)}. We will  show that our model is able to  reproduce the attenuation at a macroscopic scale, involving the combined effect of two parallel damping sources  and their crossovers.  By tuning the parameters of the model, our constitutive law  can reproduce  qualitatively and quantitatively the three regimes of acoustic attenuation versus frequency: successively $\Gamma \propto \omega^{a},\omega^4,\omega^2$ with $a \approx$ 2. 
\vg{The paper is organized as following: in section II, we introduce our viscoelastic model and derive the analytic expression of the inverse quality factor $Q^{-1}$ involving three independent parameters; in section III, based on Eq.\ref{eq:relation},  we calibrate the model on literature data of attenuation $\Gamma$ on the prototype silica glass  and demonstrate the influence of the three parameters. Finally, section IV is devoted to discussion and conclusion.
}

\section{Model} 

The material is assumed as an isotropic, homogeneous and viscous solid (amorphous materials being homogeneous and isotropic above the nanometer scales~\cite{Tsamados2009} ):  the elastic constitutive laws  can be expressed using the Hooke's law with the hydrostatic \AT{(or spherical) and} the deviatoric components 
\begin{equation}\label{eq:hooke}
    \sigma_{ij}=3K\epsilon^{sph}_{ij}+2G\epsilon^{dev}_{ij}
\end{equation}
where  $G$ is the generalized shear modulus and $K$ is the generalized bulk modulus. The spherical part of strain is  $\epsilon^{sph}_{ij}=\frac{1}{3}\delta_{ij}\epsilon_{pp}$, where $\delta_{ij}$ is Dirac function, and the deviatoric part is $\epsilon^{dev}_{ij}=\epsilon_{ij}-\frac{1}{3}\delta_{ij}\epsilon_{pp}$.

\begin{figure}[htp!]
\centering
\includegraphics[width=5cm]{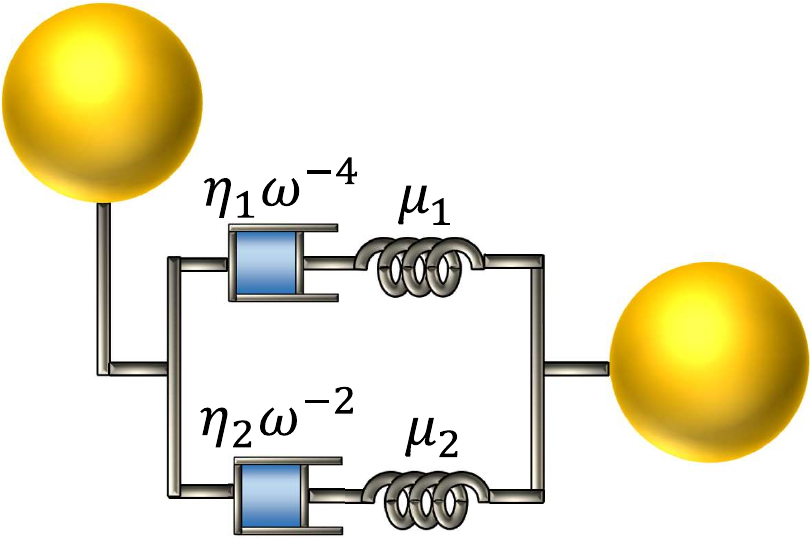}
\caption{Illustration of the viscoelastic model (PPLM): two Maxwell-like models in parallel. }\label{fig:model_parallel}
\end{figure}

In Ref.\cite{Luo2020}, we proposed to use two different simple rheological models for the spherical and deviatoric parts separately, determining the longitudinal and transverse quality factor. Here, we propose a new model for the transverse modes.
As shown in Fig.~\ref{fig:model_parallel}, \AT{in the Parallel Power-Law Model (PPLM)} we assume two Maxwell-like models in parallel, with $\mu_1$ and $\mu_2$ the rigidity and $\eta_1\omega^{-4}$ and $\eta_2\omega^{-2}$ the frequency-dependent viscosity, respectively. The physical idea behind this model is  the asynchronous activation of the  $\omega^4$ and the $\omega^2$ dependence of the attenuation,  resulting from two processes contributing additionally to the global stress in the system. 

\AT{In each Maxwell branch, the stress is related to the strain by the superposition of the elastic and of the viscous strain, that is}
\begin{equation}
\frac{d\epsilon}{dt}(t) = \frac{1}{\mu}\frac{d\sigma}{dt}(t) + \frac{1}{\eta}\sigma (t)\nonumber
\end{equation}
Using Fourier Transform, this gives 
\begin{equation}
i\omega\hat{\epsilon}(\omega)=\frac{1}{\mu}i\omega\hat{\sigma}(\omega)+\frac{1}{\eta}\hat{\sigma}(\omega)\nonumber
\end{equation}
that is, in each branch: $\hat{\sigma}(\omega)=G_0(\omega)\hat{\epsilon}(\omega)$ with
\begin{equation}
G_0(\omega)=\frac{i\omega\eta\mu}{\eta i\omega+\mu}
\end{equation}
\AT{Putting the two branches in parallel (as in Fig.~\ref{fig:model_parallel}) \vg{means} summing the stresses for an imposed deformation, therefore summing the complex moduli, hence the expression of the frequency-dependent PPLM elastic modulus $G^*(\omega)= G'(\omega) + i G''(\omega)$:}

\begin{equation}\label{eq:G_complex}
\begin{split}
G^*(\omega) &= \frac{i\mu_1\eta_1^*\omega^{-3}}{i\eta_1^*\omega^{-3} + \mu_1} + \frac{i\mu_2\eta_2^*\omega^{-1}}{i\eta_2^*\omega^{-1}+\mu_2}
\end{split}
\end{equation}

\AT{The above model can be applied, for example,  to the deviatoric components, defining a complex shear modulus which controls the dynamics of transverse waves. The full complex constitutive  tensor $\mathbb{G}^*$ can be expressed  as a  $6 \times 6$ symmetric matrix as shown in the Appendix} \vg{and  is directly related to sound propagation and attenuation.
In addition, the assumption of isotropy corresponds to the equalities  $G^*_{11}=G^*_{22}=G^*_{33}=K^*(\omega)+\frac{2}{3}G^*(\omega)$ and $G^*_{44}=G^*_{55}=G^*_{66}=G^*(\omega)$.}
From the complex constitutive elastic tensor,  longitudinal (L) and transverse (T) sound speeds can be calculated as:
\begin{align} \label{eq:sound speed}
    &\begin{cases}
   v_L(\omega)^2 &=\frac{ G'_{11}(\omega)}{\rho} \\
   v_T(\omega)^2 &= \frac{ G'_{44}(\omega)}{\rho}
    \end{cases}
\end{align}
\vg{with $\rho$ the mass density. Similarly, }the inverse quality factors ($Q^{-1}$) are obtained as:
\begin{align} \label{eq:quality factor}
    Q^{-1}_L(\omega) =\frac{ G''_{11}(\omega)}{G'_{11}(\omega)} \ \text{and} \
    Q^{-1}_T(\omega) = \frac{ G''_{44}(\omega)}{G'_{44}(\omega)}
\end{align}
where $G'_{ij}$  and $G''_{ij}$ are the reel and imaginary part of $G_{ij}$. 

In this work, we will only \AT{consider one polarization. Then, for the viscoelastic model shown in Fig.~\ref{fig:model_parallel}, we will derive the complex elastic modulus related to this polarization (see Eq.~\ref{eq:quality factor})} to get the analytic expression of the \AT{corresponding} quality factor $Q^{-1}=G''/G'$.
From Eq.\ref{eq:G_complex}, and using the power-law expression of the viscosities $\eta_1^*=\eta_1\omega^{-4}$ and $\eta_2^*=\eta_2\omega^{-2}$, $G'$ and $G''$ can be extracted:

\begin{equation}
\begin{split}
G'(\omega)&=\frac{\mu_1\eta_1^2}{\mu_1^2\omega^6+\eta_1^2} +
\frac{\mu_2\eta_2^2}{\mu_2^2\omega^2+\eta_2^2} \\
&=\mu_1(\frac{\tau_1^2}{\omega^6+\tau_1^2} + \alpha\frac{\tau_2^2}{\omega^2+\tau_2^2})
\end{split}
\end{equation}

\begin{equation}
\begin{split}
G''(\omega) &=  \frac{\mu_1^2\eta_1\omega^3}{\mu_1^2\omega^6+\eta_1^2} +
\frac{\mu_2^2\eta_2\omega}{\mu_2^2\omega^2+\eta_2^2}\\
&=\mu_1(\frac{\tau_1\omega^3}{\omega^6+\tau_1^2} + \alpha\frac{\tau_2\omega}{\omega^2+\tau_2^2})
\end{split}
\end{equation}

where $\tau_1 = \eta_1 / \mu_1 (\second^{-3})$, $\tau_2 = \eta_2 / \mu_2 (\second^{-1})$ and $\alpha = \frac{\mu_2}{\mu_1}$. As such, there are only three parameters left to characterize the attenuation.  Finally, the quality factor
 $Q^{-1}$ is given by:

\begin{equation}\label{eq:Q_new_model}
\begin{split}
Q^{-1}(\omega) =G''/G' 
&= \frac{\frac{\tau_1\omega^3}{\omega^6+\tau_1^2} + \alpha\frac{\tau_2\omega}{\omega^2+\tau_2^2}}{\frac{\tau_1^2}{\omega^6+\tau_1^2} + \alpha\frac{\tau_2^2}{\omega^2+\tau_2^2}} \\
&= \frac{\alpha\tau_2\omega^7 + \tau_1\omega^5 + \tau_1\tau_2^2\omega^3 +  \alpha\tau_1^2\tau_2\omega}{\alpha\tau_2^2\omega^6+\tau_1^2\omega^2 +(1+\alpha)\tau_1^2\tau_2^2}
\end{split}
\end{equation}

\AT{The above three independent parameters can be fitted from the sound attenuation data as shown in the next section, while the amplitude factor $\mu_1$ should be obtained from the elastic modulus of the materials. Indeed, in the low-frequency limit, it is straightforward to link $G^*(\omega)$ to  $2G$ in Eq.\ref{eq:hooke}:}

\begin{equation}
G^*(\omega \rightarrow 0) = \lim_{\omega \rightarrow 0} \frac{i\mu_1\eta_1\omega^{-3}}{i\eta_1\omega^{-3} + \mu_1} + \frac{i\mu_2\eta_2\omega^{-1}}{i\eta_2\omega^{-1}+\mu_2}  =\mu_1 + \mu_2 =  (1+\alpha)\mu_1 = 2G
\end{equation}

which leads to the following identity:

\begin{equation}
\mu_1 = 2G/(1+\alpha)
\end{equation}

\AT{Therefore, we have the frequency-dependent shear modulus $G^*(\omega)$ given in Eq.\ref{eq:G_complex} and the transverse inverse quality factor $Q^{-1}_T(\omega)$ given in Eq.\ref{eq:Q_new_model}. As mentioned  in the introduction, it is now largely accepted a $\omega^a$-$\omega^4$-$\omega^2$ dependence sequence, with $a\approx2$,  for the sound attenuation $\Gamma$.}
 \vg{ Holding the relation $  Q^{-1}  = \Gamma / \omega$, ideally, we expect to get a $\omega$-$\omega^3$-$\omega$ dependence for  $Q^{-1}$. Our model should reproduce not only such behavior, but also the amplitude of $\Gamma/ \omega$ and the crossover position.  
In the following, as detailed in the  SM Appendix, we show that our model gives a $\omega$-$\omega^3$-$\omega$ dependence for the quality factor $Q^{-1}$, and we give the analytic expression of the crossover positions. Interestingly, based on the experimental frequency inputs, we find that the temporal relation between stress and strain involves higher order temporal derivatives of the strain as compared to the Hooke's law and to the Maxwell Model, as shown in Appendix B.}
 In the next section, we will calibrate  this model on the prototype glass SiO\textsubscript{2}.

\section{Results}
\begin{figure}[htp]
\centering
\includegraphics[width=8cm]{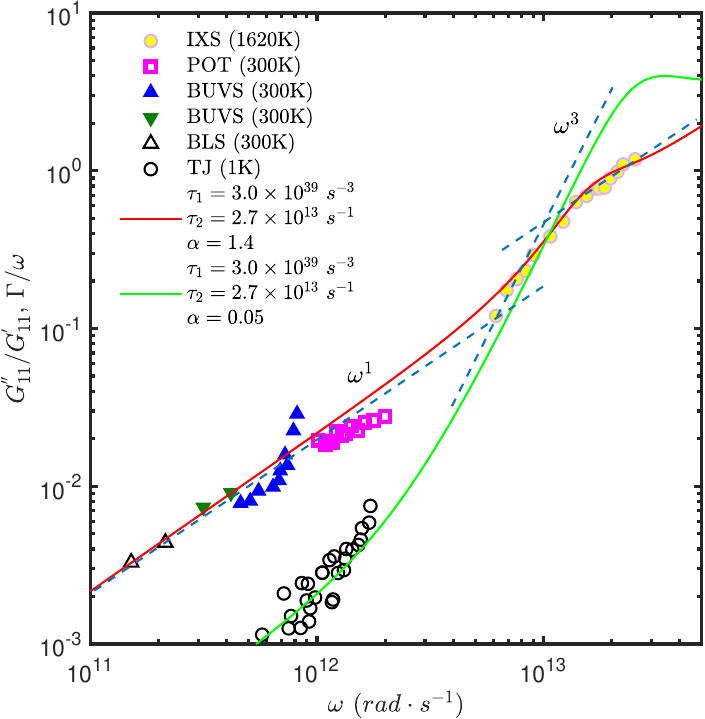}
\caption{Comparison between $\Gamma / \omega$  and $Q^{-1}$ with $\tau_1 =3\times 10^{39} \second^{-3}$, $\tau_2=2.7\times 10^{13} \second^{-1}$ and $\alpha=1.4$ (red line) or  $\alpha=0.05$ (green line) . The red (green) line is the $Q^{-1}$ of the present viscoelastic model. Experimental data are from: IXS~\cite{Baldi2010}, POT~\cite{Ayrinhac2011}, BUVS~\cite{Masciovecchio2006,Benassi2005},BLS~\cite{Vacher1980} and TJ~\cite{Dietsche1979}. The blue  dashed lines  are two power-law fits representing $\Gamma/\omega \propto \omega$ and $\Gamma/\omega \propto \omega^3$, respectively.  }\label{fig:goodfit}
\end{figure}

In order to use our model, we need to calibrate it against attenuation results from microscopic experimental or theoretical results. Similarly to what we have done in Ref.\cite{Luo2019}, we calibrate our model on literature data on the prototype glass SiO\textsubscript{2}. Reported attenuations are for longitudinal modes only, but it has been demonstrated that the same phenomenology is present in transverse modes at higher values of $q$ and $\omega$, making the computations more comfortable~\cite{Mizuno2020}. 
The experimental results of SiO\textsubscript{2} are reported in Tab.~\ref{tab:sound_attenuation_literatre_1} and in Fig.~\ref{fig:goodfit}
in which sound attenuation  $\Gamma$ is measured at 300K in the subterahertz range ($\nu < 0.3$ THz with  $\omega=2\pi \nu$) and at 1620K at terahertz frequencies ( $\nu> 1$ THz).  \AT{Two sets of data can be identified: a set of high temperature data ($T> 300 K$) and a concurrent set of data at $T=10 K$ in the low frequency range.} The three regimes mentioned before can be clearly found in the figures. At high frequency, a  $\omega^4$-$\omega^2$  crossover of $\Gamma$ appears around $\omega = 1 \times 10^{13} \ \rad \cdot \second^{-1}$ ($\nu= $ 1.5 THz)  as shown in Fig.~\ref{fig:goodfit}. In addition, in the subterahertz range,  there should be another $\omega^2$-$\omega^4$ crossover which is not explicitly given by the experimental data but shown as  the intersection of  the auxiliary lines ($\omega^2$ and $\omega^4$). 
\vg{In Fig.~\ref{fig:goodfit} we report our best fit for high (red line) and low (green line) temperature, obtained adjusting the parameters of the model $\tau_1$, $\tau_2$ and $\alpha$, as explained later. The fitting parameters for both temperature ranges are reported in Tab.~\ref{tab:para_2}. The results of the fit at high temperature is excellent over the full frequency range.} 

\AT{For the low temperature data, the departure from the high temperature data is very clear at low frequencies and can be captured with the same model but with another set of parameters as will be discussed later (see Table~\ref{tab:para_2}). } 
 
 \begin{table*}[htp]
\caption{Experimental results of sound attenuation of glasses in the literature. (TJ: tunneling junction;
 BUVS: Brillouin  ultraviolet  light  scattering;
 IUVS: inelastic ultraviolet scattering;
 POT: picosecond optical technique;
 IXS: inelastic X-rays scattering;
 BLS: Brillouin light scattering.
 ) }
\centering
\begin{threeparttable}
\begin{tabular} {|c|c|c|c|c|c|c|}
\hline
 \thead{Materials} &\thead{Method} & \thead{T (K)} & \thead{Power}& \thead{Frequency range  and/or $\nu_c$}  & \thead{LA velocity (m/s)} &\thead{Ref} \\ \hline
 \multirow{7}{*}{v-SiO\textsubscript{2} } & TJ & 1  & 4 & 0.1-0.4 THz  &  & \cite{Dietsche1979} \\ \cline{2-7}
 &  BUVS  & 300 &2  & \makecell{5-70 GHz\tnote{*}}   &  & \cite{Benassi2005} \\ \cline{2-7}
&  IUVS & 300 & 2 & 75-95 GHz\tnote{*}     &  & \cite{Masciovecchio2004} \\ \cline{2-7}
& IUVS  & 300 & 2-4 & \makecell{75-150 GHz\\$\nu_c=100 \pm 10$ GHz}   & 5950 &\cite{Masciovecchio2006} \\ \cline{2-7}
& POT   & 300 & 2 & \makecell{near 250 GHz}   & 5940 & \cite{Devos2008} \\ \cline{2-7}
& IXS  & 1050 & 2 & \makecell{1-3 THz}   & 5800 &\cite{Benassi1996} \\ \cline{2-7} 
& IXS &1620  & 4-2 &  \makecell{1- 4THz\\$\nu_c=$1.5 THz}  & 6500 & \cite{Baldi2010} \\ \hline 
d-SiO\textsubscript{2} &  IXS& 565 & 4.21 & 1.2-1.9 THz &  & \cite{Ruffle2003}\\ \hline
 \multirow{2}{*}{\makecell{\HL{Li\textsubscript{2}O-2B\textsubscript{2}O\textsubscript{3}}}  }  & BLS  & \makecell{ 300 \\573} & 1 & 20-40 GHz & 7600 & \cite{Ruffle2006} \\ \cline{2-7}
  & IXS & 573 & 4 & 1.2-2.4 THz  &  & \cite{Ruffle2006} \\ \hline
 \multirow{2}{*}{ \makecell{ Glycerol } } 
  & IXS  & \makecell{ 16\\167} & 2 & 0.6-3.5 THz \tnote{$\dagger$}    &  & \cite{Ruocco1999} \\  \cline{2-7} 
 & IXS  & 150.1 & 4-2 & \makecell{1-2.5 THz \\ $\nu_c =$1.2 THz}   & 3630 &\cite{Monaco2009} \\ \hline
    \makecell{BeF\textsubscript{2}}& IXS & 297 & 2 & \makecell{0.6-6 THz}   & 5500  &   \makecell{ \cite{Scopigno2003}\\\cite{Buchenau2014} }\\ \hline
    \makecell{Poly butadiene}&  IXS & 140 & 2 & \makecell{0.6-3.6 THz}    &  2770 &\makecell{\cite{Fioretto1999}\\\cite{Buchenau2014} } 
 \\ \hline
\end{tabular}
  \begin{tablenotes}
    \item[*] Use the sound speed value in Ref.\cite{Masciovecchio2006}.
    \item[$\dagger$] Use the sound speed value in Ref.\cite{Monaco2009}.
  \end{tablenotes}
\end{threeparttable}
\label{tab:sound_attenuation_literatre_1}
\end{table*} 

\begin{table}[htp!]
\caption{Parameters of the new viscoelastic continuum model PPLM. Calibration on silica glass at low ($T=1K$)  high temperatures ($T > 300 K$)~\cite{Ayrinhac2011,Baldi2010}.}
\centering
\begin{tabular}{|c|c|c|c|}
\hline
$T$ & $\tau_1 = \eta_1/\mu_1 (\second^{-3})$ & $\tau_2 = \eta_2/\mu_2 (\second^{-1})$ & $\alpha = \mu_2/\mu_1 $ \\
\hline
$1 K$ & $3\times 10^{39}$ & $2.7\times 10^{13}$ &  0.05  \\ 
\hline
$> 300 K$ & $3\times 10^{39}$ & $2.7\times 10^{13}$ &  1.4  \\
\hline
\end{tabular}
\label{tab:para_2}
\end{table}

It is worth going more in the detail of the fitting procedure to highlight the role of each parameter.
A good fit should give a $Q^{-1}$ reproducing not only  the amplitude of the acoustic attenuation, but also at least the crossover from $\omega^4$ to $\omega^2$ around $1 \times 10^{13}~ \rad \times \second^{-1}$ ($\nu= \omega/2\pi=$ 1.5 THz). It is not easy to reach these two conditions through a simple  optimization algorithm, because most of the time it only meets the amplitude  condition but the crossover condition is ignored.  To meet the two conditions simultaneously, we start with a know-how to illustrate the evolution  of the amplitude and the crossover frequencies versus  the three parameters  $\tau_1$, $\tau_2$ and $\alpha$.  This approach is necessary as the first step  of the parameters identification before finding an alternative optimization algorithm in  future work.  

\begin{figure}[htp!]
\centering
\includegraphics[width=7cm]{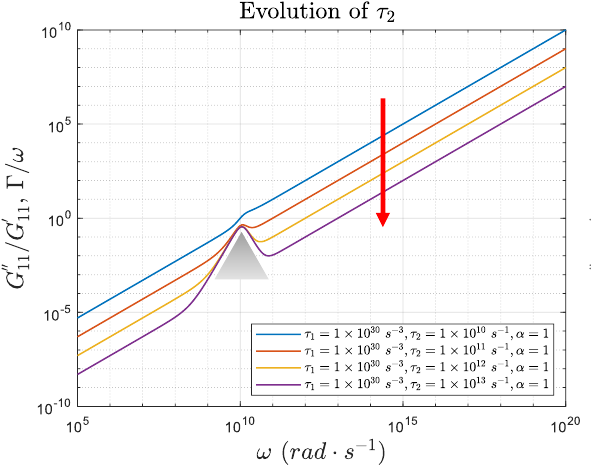}\\
\includegraphics[width=7cm]{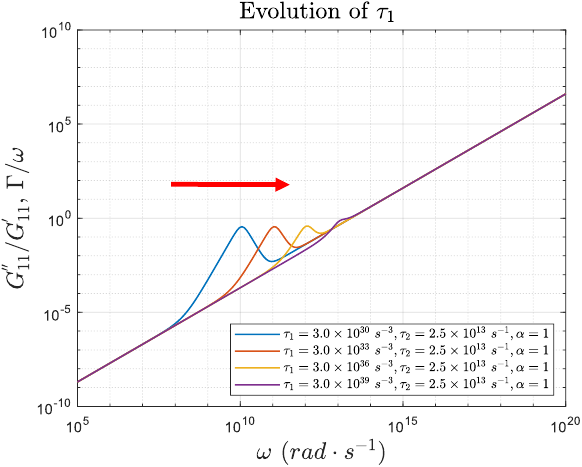}\\
\includegraphics[width=7cm]{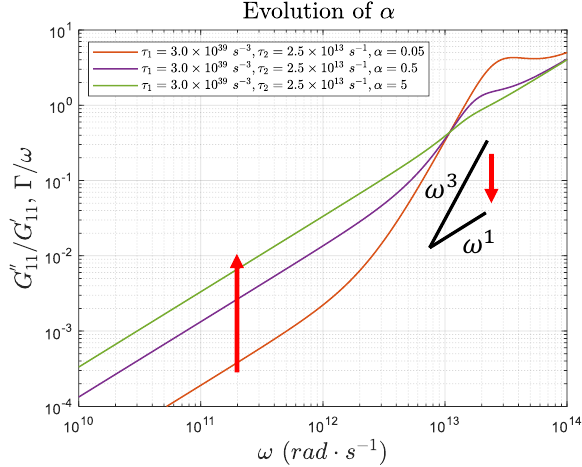}\\
\caption{Evolution of $Q^{-1}$ as a function of $\tau_2$ (top), $\tau_1$ (middle) and $\alpha$ (bottom), respectively. The red arrow indicates the growth direction of the parameter.}\label{fig:parameters}
\end{figure}

The roles of the three parameters are  displayed one by one  in Fig.~\ref{fig:parameters}. By default, we set $\tau_1= 10^{30} ~ \second^{-3},\tau_2=10^{10}~ \second^{-1}$ and $\alpha=1$. The chosen order of magnitude for different parameters can be estimated  based on the crossover positions.

Let's first consider  $\tau_2$, as shown in the left panel of Fig.~\ref{fig:parameters}. We can find that the global amplitude of the $Q^{-1}$ decreases with $\tau_2$. Therefore, in the first step, we change $\tau_2$ in order to get the same order of magnitude for $Q^{-1}$ obtained in the numerical model, and the experimental values of $\Gamma/\omega$.  During this process, one peak appears at $\omega=1\times 10^{10}  ~ \rad \cdot \second^{-1}$ whose position and amplitude are determined by  the two  other parameters. This peak corresponds to a transition between two different dominant frequency powers in the constitutive model and depends on the relation among three parameters of the model. In the current configuration, that is for $\tau_1=1\times 10^{30} ~  \second^{-3} $ and $\alpha =1$, it shows a $\omega^3-\omega^{-3}$ crossover.

Let's now consider the variations of $\tau_1$, after having optimized $\tau_2$. The influence of $\tau_1$ is shown in the middle panel of Fig.~\ref{fig:parameters}. Increasing $\tau_1$ pushes to the right the above $\omega^3$-$\omega^{-3}$ peak. When this peak moves to the right, it becomes smaller and the $\omega^{-3}$ part disappears first due to the collapse with the increasing $\omega$ background.   In this step, only the peak moves and the amplitude of left or right wing is unchanged.

Finally, let's consider the role played by the ratio $\alpha=\mu_1/\mu_2$, which characterizes the rigidity ratio  between the two parallel processes of dissipation contributing additionally to the stress state. As shown in the right panel of Fig.~\ref{fig:parameters}, changing $\alpha$ will raise or lower the left wing and modify the power law near the  crossover between left and right wing. When $\alpha \ll 1$, the left wing will  drop significantly. However, when $\alpha \gg 1$, the crossover $\omega^3$-$\omega^{-3}$ will progressively disappear and the whole line, especially the part $\propto\omega^3$, tends to a single $\omega$ power law.  Between them, the $\omega^3$ tends  to $\omega$. Therefore, an appropriate $\alpha$ should raise the left wing  to fit the experiments data, while keeping the $\omega^3$ apparent power law as much as possible.

In addition, the relaxation times  $\tau_1 \omega^{-4}$ and $\tau_2\omega^{-2}$ should make sense physically, which requires them being longer than $1\times 10^{-14}  - 1\times 10^{-15}  ~ \second$.  Using the orders of magnitude of $\tau_1$ and $\tau_2$, we estimate the corresponding relaxation time in Tab.\ref{tab:ch5_relaxation_time}. Relaxation times shorten as the frequency increases, indicating as upper frequency limit for this model $3 \times 10^{13}$ Hz, that is satisfactory.

\begin{table}
\caption{Estimation of the relaxation time for $\tau_1$ = $3\times 10^{39}$  $\second^{-3}$ and $\tau_2$ = $2.7\times 10^{13}$   $\second^{-1}$.}\label{tab:ch5_relaxation_time}
\centering
\begin{tabular}{|c|c|c|}
\hline
$\omega$ ($\rad \cdot \second^{-1}$) & $\tau_1 \omega^{-4}$  (s) & $\tau_2 \omega^{-2}$ (s)\\ \hline
$1\times 10^{11}$ & $3\times 10^{-5}$ & $2.7\times 10^{-9}$  \\
$1\times 10^{12}$ & $3\times 10^{-9}$ & $2.7\times 10^{-11}$  \\
$1\times 10^{13}$ & $3\times 10^{-13}$ & $2.7\times 10^{-13}$ \\
$3\times 10^{13}$ & $3.7\times 10^{-15}$ & $3\times 10^{-14}$ \\
\hline
\end{tabular}
\end{table}

In SM Appendix, we discuss in more details the conditions giving rise to a $\omega-\omega^3-\omega$ behavior. If these conditions are strictly met, then the crossover frequencies can be expressed in terms of $\alpha$, $\tau_1$ and $\tau_2$. We recall the two crossover frequencies evidenced in SM Appendix:

\begin{itemize}
\item Position of the crossover $\omega$ - $\omega^3$:
\begin{equation}\label{eq:position1}
\omega_{1-3} = \frac{1}{\alpha+1} \tau_2
\end{equation}

\item Position of the crossover $\omega^3$ - $\omega$:
\begin{equation}\label{eq:position2}
\omega_{3-1} = (\frac{1}{\alpha})^{1/2}(\frac{1}{\alpha+1})^{1/2} \tau_2
\end{equation}
\end{itemize}

Note that in the present model, the ratio $\omega_{3-1}/\omega_{1-3}$ depends only on $\alpha$. It is given by :
\begin{equation}
\frac{\omega_{3-1}}{\omega_{1-3}} = \sqrt{\frac{\alpha+1}{\alpha}}
\end{equation}
where $\alpha > 0.76$ is a necessary condition to have a perfect $\omega-\omega^3-\omega$ behavior without exhibiting other power laws (this condition disappears if other crossovers are allowed). This gives an upper limit of the ratio $\omega_{3-1} / \omega_{1-3}$ less than 1.52. 
In practice, these conditions may  not be strictly met but were sufficient in our case. Moreover, the results about the crossover frequencies  are still very useful because they  determine the magnitude of the parameters. For example,  knowing that $\omega_{3-1} \approx 2\pi\times 1.5$ THz $\approx 9.4 \times 10^{12} ~ \rad \cdot \second^{-1} $  and $\alpha$ is normally supposed as 1 at first, then we have $\tau_2=\omega_{3-1} / \sqrt{0.5} \approx 1.3 \times 10^{13}  ~  \second^{-1}$ and $\tau_1 = \frac{\tau_2^3}{4} \approx 5  \times 10^{38} ~  \second^{-3}$.\AT{ It is then possible to get a coherent set of parameters in the PPLM model reproducing the experimental data of acoustic attenuation in  silica glass. The data for silica at $T>300K$ and $T=10K$ are shown in Table~\ref{tab:para_2} and will be discussed in the next part.}

\section{Discussion and Conclusion}

Our model is a first but successful attempt to find a continuous model taking into account dissipative processes at small scales. It is based on exact solutions known at a microscopic level for the apparent acoustic attenuation in glasses. The microscopic \vg{behavior is} taken into account through the frequency dependence of the viscosity. The model does not need the exact description of structural disorder at the atomic scale but acts as a reduced model for the dynamical response of a glass. As such, it is not simply a data driven model, but it includes atomistic relaxation times and interactions as hidden variables. Concerning the parameters included in the model, we have shown that it is possible to find  parameters allowing a good description of experimental data on a large range of frequencies. Following these fits, thermal effects are shown to act through the ratio between harmonic and anharmonic contributions to acoustic attenuation, driven by the apparent rigidity of the two processes. Our model is valid only up to the Ioffe-Regel frequency, since above this frequency, dynamical equations involving linear visco-elastic models with sound propagation and attenuation, must be replaced by diffusive equations, in agreement with the high limit values of the characteristic times invlovled in our model (see Tab.~\ref{tab:ch5_relaxation_time}). It has been shown in another article that, at higher frequencies, attenuation becomes temperature independent and is only controlled by the local structural symmetries~\cite{Damart2017} thus giving rise to \vg{another frequency dependence}. But these very high frequencies are hardly reachable experimentally. Our model is thus valid to describe waves propagation and attenuation at all the frequencies included in the ballistic regime of phonons.

\vg{In order to reproduce multiple power-law frequency dependencies of the viscosity in glasses, our viscoelastic model is based on two viscosities, with different power-law exponent, acting in parallel, and it has been shown to correctly reproduce the frequency dependence of acoustic attenuation in silica glass at different temperatures. }The original idea is to relate the global behaviour of the glass at a continuous scale, to microscopic processes: acoustic scattering for the first one and anharmonicity for the second one. Due to the large distribution of energy barriers in glasses, these two processes act in parallel, with coexistence of elastic processes ($1/\eta_1^*\propto\omega^4$) and irreversible processes ($1/\eta_2^*\propto\omega^2$). The assumed frequency dependence of the viscosity is inversely proportional to the frequency dependence of the relaxation times (or attenuation times) of these two different processes, as already been highlighted in glasses~\cite{Gelin2016,Mizuno2020}. The characteristic times involved in our model are realistic. The microscopic origin of these different processes is thus very clear, but the relative weight of each of these processes is not so obvious, due to the difficulty of identifying and counting all related events at a microscopic level. Its temperature dependence is thus taken into account empirically by comparing the data to the experimental results. It shows that the very low temperature range ($T=1K$) must be handled apart from higher temperatures ($T>300K$), these latter been generally associated. Temperature acts here as a stiffener for the anharmonic process (or a softener for the harmonic one), thus increasing the weight of the anharmonic process in the internal friction as compared to the one of disorder induced harmonic acoustic scattering. Our model is not only able to \vg{reproduce} all the accessible experimental data, but \vg{also allows their reintrepretation}. For example, the "cross-over" identified in~\cite{Masciovecchio2006} is \vg{likely not correct and within the error bars,  as also supported by the very good alignment of higher frequency data with our model. Concerning} the low temperature case~\cite{Dietsche1979} the data are very well described as an extended cross-over between the $\omega$ and the $\omega^3$ regime of internal friction. \vg{There is now a need for more experimental data in intermediate temperature ranges to definitely confirm our model. In the low temperature case, it would be extremely interesting to compare the model with data in the 0.1-10 THz range. Unfortunately, no data in this range at low temperature are available in literature, thus a single change in $\alpha$ appears to be sufficient here to switch from the high temperature to the low temperature case. The sensitivity of the moduli to the temperature have also been discussed by other authors in the context of the glass transition~\cite{Dyre2006}}

\vg{We have also applied the PPLM model to other glasses, with different chemical composition, as listed in Tab.\ref{tab:sound_attenuation_literatre_1}. In the case of glycerol,  its sound attenuation is evidently weaker than that of silica glass, indicating a  larger $\tau_2$ (Fig.\ref{fig:parameters}). Then, it is reported a $\omega^2 - \omega^4$ crossover around $\nu_c =$1.2 THz at 150.1K \cite{Monaco2009}, which is comparable to that of silica glass ($\nu_c =$1.5 THz). As such, we may expect a larger $\alpha$ value (Eq.\ref{eq:position2}). Extracting the data from Ref\cite{Monaco2009}, we could fairly reproduce them using $\tau_1=3e39$, $\tau_2=7e14$ and $\alpha=30$, confirming our  expectation.
A special case is represented by  Li\textsubscript{2}O-2B\textsubscript{2}O\textsubscript{3}, which presents a $\omega - \omega^4$ crossover  at 573K \cite{Ruffle2006}. We have found that the generalization of the PPLM model can reproduce this crossover by defining $\alpha_2=1$, as reported in Appendix.3. As for  other materials that only have a $\omega^2$ behavior, the Maxwell model is enough. }

We have based our approach on the frequency dependence of the viscosity, since it is the best way for comparing the resulting behaviour to the available experimental data. Still, it is interesting to relate it as well to the behaviour in the time domain. For that, an inverse Fourier Transform of the constitutive laws is needed, as detailed in Appendix B. The calculation of the temporal behaviour based on this approximative description of the frequency dependence of the parameters shows two interesting characteristics: first a departure from the initial Maxwell Model with the emergence of a Kelvin-like model, second the contribution of higher order derivatives (first and third order derivatives of the stress related to the strain and its second order derivatives), induced by memory effects. Note that such kind of power-law viscosity has already been used to describe more accurately biological samples as well~\cite{Perez2015}.

Provided the parameters $\tau_1$, $\tau_2$ and $\alpha$, the interesting case with  three principal frequency dependencies of the internal friction $\omega$-$\omega^3$-$\omega$   is  only a special case  offered by the  visco-elastic model.  
For this special case, we can write down the asymptotic approximation for the three frequency dependencies : (1) Low frequency: $Q^{-1}_{(1)} \sim \frac{\alpha}{1+\alpha} \frac{1}{\tau_2}\omega$; (2) Intermediate frequency: $Q^{-1}_{(2)} \sim \frac{1}{1+\alpha} \frac{1}{\tau_1}\omega^3$; (3) High frequency: $Q^{-1}_{(3)} \sim \frac{1}{\tau_2}\omega$. It is noticed that the process with viscosity $\eta_2 \omega^{-2}$ dominates, even if for different reasons, both low  and high frequency regimes,  while the $\eta_1 \omega^{-4}$ process activates mainly at intermediate frequencies corresponding to the Rayleigh-like damping below the Ioffe-Regel frequency. Especially, it can be seen that,  when $\alpha \gg 1$, $Q^{-1}_{(1)}$ tends to $\frac{1}{\tau_2}\omega$, thus $Q^{-1}_{(3)}$,  \vg{and the intermediate regime  disappears}, as illustrated in Fig.\ref{fig:parameters}. In contrast, when $\alpha \ll 1$, the low-frequency $\omega$ process gives a smaller value of the quality factor compared to the high-frequency ones. This $\alpha$-dependent $\omega$ process is impressively adapted to describe the temperature-dependent acoustic attenuation $\Gamma \propto \omega^2$ ($G''/G'\propto\omega$) at low frequencies (see Fig.\ref{fig:goodfit}). This low frequency behavior is known to be due to anharmonicity\AT{, for example induced by thermal activation and often described with the help of two level systems~\cite{Phillips1987}, as} also shown in MD simulations \cite{Mizuno2020}. As such, small $\alpha$ corresponds to low temperature and large $\alpha$ corresponds to a relatively high temperature, meaning that the effective role of the temperature would be to \AT{stiffen the $\eta_2\omega^{-2}$ process by increasing its rigidity $\mu_2$ compared to $\mu_1$}. Moreover, our model predicts an extension of the $\omega^3$ regime \AT{of the internal friction} at high frequencies in the low temperature case as can be seen in Fig.\ref{fig:goodfit} and possibly additional frequency dependencies at higher frequencies that are difficult to reach experimentally. As can be inferred from Appendix C., in the low frequency regime, the expression of $Q^{-1}_{(1)}$ results from the combination of the viscous terms in the two parallel contributions to the stress, while the results obtained for $Q^{-1}_{(3)}$ \vg{come from the combination of the elastic response of the harmonic process with the viscous term of the anharmonic process}. Note finally that, as shown in  Appendix C, if the anharmonic process is related to another power-law dependent viscosity $\eta_2^*=\eta_1\omega^{\alpha_2}$, then the low frequency behaviour of the internal friction becomes $Q^{-1}_{(1)} \sim \frac{\alpha}{1+\alpha} \frac{1}{\tau_2}\omega^{-\alpha_2-1}$ as soon as $\alpha_1<\alpha_2\leq -1$, and $Q^{-1}_{(3)} \sim \frac{1}{\tau_2}\omega^{-\alpha_2-1}$ as well, for the same conditions ($\alpha_1$ being the power exponent of the harmonic process). This means that, whatever the precise value for the power-law of the anharmonic process, this power-law will dominate \vg{both low-frequency and high frequency responses}. The complete calculation, reported in Appendix C, shows that the intermediate regime is sensitive to the other dissipative process. This is not in contradiction with the experimental data (see Tab.\ref{tab:sound_attenuation_literatre_1}): it reflects perfectly the ambiguous value of the exponent measured at low frequency, while the high frequency regime is difficult to reach experimentally due to the Ioffe-Regel limit. \AT{In the low frequency regime, different interpretations are competing for the exat value of the exponent: either an enhanced sensitivity to quenched stresses~\cite{Gelin2016}, either a frequency dependence of the mean-free path $\propto\omega$ due to the Fermi golden rule between separated energy levels~\cite{Phillips1987}, either scattering on irreversible defects like Eshelby inclusions~\cite{Churochkin2016}.} More importantly, the low and the intermediate regimes are well preserved. Finally, as shown in the Tab.\ref{tab:ch5_relaxation_time}, the frequency-dependent relaxation times given by the fitted values of three parameters are reasonable up to $\omega=$ $30 \times 10^{12}~ \rad \cdot  \second^{-1}$ which covers all spectral range of experimental data. Therefore, we can conclude that the values of this model are physically meaningful. 

In conclusion, the  model presented here, which involves two parallel sources of dissipation with a frequency dependent viscosity, \vg{is shown to be able to reproduce the successive  acoustic attenuation regimes $\omega^2$ - $\omega^4$ - $\omega^2$ for the internal friction, or equivalently, $\omega$ - $\omega^3$ - $\omega$ regimes for the quality factor}, observed in glasses.  In practice, we have presented a calibration of the model on a-SiO\textsubscript{2} which gives a quality factor $Q^{-1}$ that fits very well the experimental attenuation data $\Gamma / \omega$  \cite{Baldi2010,Ruffle2003,Ruocco1999,Masciovecchio2004,Benassi2005,Masciovecchio2006,Devos2008,Dietsche1979,Benassi1996}. 
Interestingly, the frequency dependence of the viscosities involves two complementary mechanisms: one is related to an attenuation time $\tau_1 \omega^{-4}$ (possibly related to low acoustic scattering at the atomic scale as discussed before  \cite{Gelin2016}), while the other involves a relaxation time $\tau_2 \omega^{-a}$ with $a\approx 2$ (possibly related to anharmonicity or to strong acoustic scattering at small scale, as already discussed in the introduction). The permanent combination of these two well-known atomistic processes in parallel (additional contribution to the stress for a given strain) allows to reproduce the two experimentally observed crossovers in the frequency dependence of the attenuation. The comparison of our model with experimental data on a large frequency range leads to the determination of the 3 parameters involved in the model, that is: $\tau_1 = 3 \times 10^{39} ~ \second^{-3}$, $\tau_2=2.7\times 10^{13}   ~ \second^{-1}$ and $\alpha = 1.4$ at ambient temperature (300K) but is also able to reproduce the temperature dependence in the low frequency part, by tuning only the parameter $\alpha$, which is lowered to $0.05$ at 1 K. These numerical values are reasonable for $\omega < 3 \times 10^{13}~ \rad \cdot \second^{-1}$ ,that is the largest frequency that can be reached at the atomic scale. Thus, our results on one hand give a description of the effective frequency-dependent acoustic attenuation in SiO\textsubscript{2} glass in a very large frequency range, spanning three orders of magnitude from GHz to THz, and on the other hand consolidate the  possibility of  homogenizing the multi-channel
intrinsic acoustic attenuation from nanometric to macroscopic scale with appropriate viscoelastic models. 
As such, this model will allow for improved and more realistic large scale simulations of metamaterials with an amorphous component. As an example, in a recent work, we have used continuum finite element 
modeling to study the acoustic attenuation in nanocomposites, which has allowed us to study the effect of the interface
scattering on the real size materials, underling the existence of different transport regimes (propagative, diffusive, localized and mixed regime), depending on the nanostructuration lengthscale and elastic mismatch between the components. However, in that work, the component materials are purely elastic and non-dissipative, so the intrinsic sound attenuation is not considered  \cite{Luo2019}. The introduction of the dissipation in the amorphous matrix would allow to evidence the competition between extrinsic and intrinsic scattering sources, and to predict the efficiency of the nanostructuration in attenuating phonons in a medium where intrinsic attenuation is  important. Of course, we expect to be confronted to different regimes, depending on phonon frequency: a key role of the extrinsic source for low intrinsic scattering, and a less important effect in the strong scattering regime. \vg{Similarly, this type of complex dissipative constitutive law will allow taking properly into account thermo-mechanical couplings when they exist. This will allow a more realistic description of the large scale effect of competing dissipative processes initiated at a very small scale. Other applications include: high-frequency engineering, and shock waves attenuation. The application to the calculation of the thermal conductivity in the ballistic regime yields not only the calculation of the attenuation length, but additionally a proper description of the vibrational density of states, including acoustic wave dispersion effects (frequency dependence of the sound wave or equivalently frequency dependence of the effective elastic constante). }

Finally, our model paves the way to further studies, investigating other power laws for phonon attenuation, introducing anisotropy and studying the effect of temperature, allowing to describe acoustic propagation  as well as thermal  transport in structured materials of arbitrary complexity.

\section*{Appendix A: 3D complex constitutive tensor}
\HL{
Constitutive tensor, stress and strain  can be expressed in matrix form. In three dimensional case, due to the symmetry of shear strain $\epsilon_{ij}=\epsilon_{ji}$ and shear stress $\sigma_{ij}=\sigma_{ji}$, 
strain and stress vector are reduced to 6 terms and read  respectively in Voigt notation :
\begin{center}
$\epsilon_{ij} =
\begin{bmatrix}
    \epsilon_{11}\\ \epsilon_{22}\\ \epsilon_{33}\\2\epsilon_{12}\\2\epsilon_{13}\\ 2\epsilon_{23}
\end{bmatrix} $ 
and
$\sigma_{ij} =
\begin{bmatrix}
    \sigma_{11}\\ \sigma_{22}\\ \sigma_{33}\\\sigma_{12}\\\sigma_{13}\\ \sigma_{23}
\end{bmatrix} $
\end{center}
It is conventional to express the shear strain as $2\epsilon_{ij} = \gamma_{ij}$ which is called engineering shear strain.
}

\HL{
The symmetric constitutive tensor $\mathbb{G}$  ($G_{ij}=G_{ji}$) can be developed as :
\begin{equation}
\mathbb{G}=
\begin{bmatrix}
    G_{11}&G_{12}&G_{13}&G_{14}&G_{15}&G_{16}\\
    G_{21}&G_{22}&G_{23}&G_{24}&G_{25}&G_{26}\\
    G_{31}&G_{32}&G_{33}&G_{34}&G_{35}&G_{36}\\
    G_{41}&G_{42}&G_{43}&G_{44}&G_{45}&G_{46}\\
    G_{51}&G_{52}&G_{53}&G_{54}&G_{55}&G_{56}\\
    G_{61}&G_{62}&G_{63}&G_{64}&G_{65}&G_{66}\\
\end{bmatrix} 
\end{equation}
}

\section*{Appendix B: Constitutive law in time domain}
\AT{{\bf A.1. Maxwell Model}}
\AT{The Maxwell Model follows the equation of evolution
\begin{equation}
\frac{d\epsilon}{dt}(t) = \frac{1}{\mu}\frac{d\sigma}{dt}(t) + \frac{1}{\eta}\sigma (t)\nonumber
\end{equation}
Using Fourier Transform $\hat{\epsilon}(\omega)=\int_{-\infty}^{+\infty}\epsilon(t)\exp{-i\omega t}\, dt$, this gives 
\begin{equation}
i\omega\hat{\epsilon}(\omega)=\frac{1}{\mu}i\omega\hat{\sigma}(\omega)+\frac{1}{\eta}\hat{\sigma}(\omega)\nonumber
\end{equation}
thus yielding to
\begin{eqnarray}
\hat{\sigma}(\omega)&=&\frac{i\omega\eta\mu}{\eta i\omega+\mu}\hat{\epsilon}(\omega)\nonumber\\
&=&\left( \frac{\eta^2\mu\omega^2}{\eta^2\omega^2+\mu^2}+i\frac{\eta\mu^2\omega}{\eta^2\omega^2+\mu^2}\right)\hat{\epsilon}(\omega)
\label{eq:MaxwellAppB}
\end{eqnarray}
whose inverse Fourier Transform allows getting, using $TF\left(\frac{1}{2\eta}\exp{-\frac{\mu}{\eta}\vert t\vert}\right)=\frac{\mu}{\eta^2\omega^2+\mu^2}$ combined with two integrations by part, the well known real-time stress:
\begin{equation}
\sigma(t)=\frac{1}{2\pi}\int_{-\infty}^{+\infty}\hat{\sigma}(\omega)\exp{+i\omega t}\,d\omega = \mu\int_{-\infty}^t\dot{\epsilon}(u)\exp{\left(-\frac{\mu}{\eta}(t-u)\right)}\,du\nonumber
\end{equation}
}

\AT{{\bf A.2 Parallel Power-Law Model}}

\AT{In this case, 
\begin{equation}
\hat{\sigma}=\hat{\sigma}_1+\hat{\sigma}_2=\lbrack G_1'(\omega)+G_2'(\omega)+i\left(G_1"(\omega)+G_2"(\omega)\right)\rbrack\hat{\epsilon}(\omega)
\end{equation}
Putting $\eta_1^*=\eta_1\omega^{-4}$ and $\eta_2^*=\eta_2\omega^{-2}$ in Eq.~\ref{eq:MaxwellAppB} and noting $\tau_1=\eta_1/\mu_1$ and $\tau_2=\eta_2/\mu_2$, we have:
\begin{eqnarray}
G_1'(\omega)=\frac{\mu_1\tau_1^2}{\omega^6+\tau_1^2}\nonumber\\
G_2'(\omega)=\frac{\mu_2\tau_2^2}{\omega^2+\tau_2^2}\nonumber\\
G_1"(\omega)=\frac{\tau_1\mu_1\omega^3}{\omega^6+\tau_1^2}\nonumber\\
G_2"(\omega)=\frac{\tau_2\mu_2\omega}{\omega^2+\tau_2^2}\nonumber
\end{eqnarray}
Let's consider first the {\bf harmonic model} $G_1(\omega)$. The inverse Fourier Transform allows getting the real-time stress:
\begin{equation}
\sigma_1(t)=\frac{1}{2\pi}\int_{-\infty}^{+\infty}\left\lbrack\frac{\mu_1\tau_1^2}{\omega^6+\tau_1^2}\hat{\epsilon}(\omega)+\frac{\tau_1\mu_1}{\omega^6+\tau_1^2}i\omega^3\hat{\epsilon}(\omega)\right\rbrack\exp{(i\omega t)}\,d\omega
\end{equation}
This can be solved using the elementary decomposition:
\begin{eqnarray}
\frac{1}{\omega^6+\tau_1^2}&=&\frac{1}{6\alpha^5}\left(\frac{1}{\omega-\alpha}-\frac{1}{\omega+\alpha}\right)+\frac{e^{i\frac{\pi}{3}}}{6\alpha^5}\left(\frac{1}{\omega-\alpha e^{i\frac{\pi}{3}}}-\frac{1}{\omega+\alpha e^{i\frac{\pi}{3}}}\right)\nonumber\\
&+&\frac{e^{-i\frac{\pi}{3}}}{6\alpha^5}\left(\frac{1}{\omega-\alpha e^{-i\frac{\pi}{3}}}-\frac{1}{\omega+\alpha e^{-i\frac{\pi}{3}}}\right)
\end{eqnarray}
with $\alpha=\tau_1^{1/3}\exp{(i\frac{\pi}{6})}$, combined with 
\begin{equation}
TF^{-1}\left(\frac{1}{\omega-\alpha}\right)=i e^{i\alpha t}.sign\left(Im(\alpha)\right).H\lbrack t.sign\left(Im(\alpha)\right)\rbrack
\end{equation}
for $Im(\alpha)\ne 0$. We thus get:
\begin{eqnarray}
\sigma_1(t)&=&\frac{\mu_1\tau_1^2}{6\tau_1^{5/3}}\int_{-\infty}^{+\infty}\left\lbrack e^{(-\tau_1^{1/3}\vert t-u\vert)}+2e^{(-\frac{\tau_1^{1/3}}{2}\vert t-u\vert)}\sin\left(\frac{\tau_1^{1/3}\sqrt{3}}{2}\vert t-u\vert + \frac{\pi}{6}\right)\right\rbrack\epsilon (u)\, du\nonumber\\
&-&\frac{\mu_1\tau_1}{6\tau_1^{5/3}}\int_{-\infty}^{+\infty}\left\lbrack e^{(-\tau_1^{1/3}\vert t-u\vert)}+2e^{(-\frac{\tau_1^{1/3}}{2}\vert t-u\vert)}\sin\left(\frac{\tau_1^{1/3}\sqrt{3}}{2}\vert t-u\vert + \frac{\pi}{6}\right)\right\rbrack\frac{d^3\epsilon}{du^3} (u)\, du\nonumber
\end{eqnarray}
that can be rewritten, after few integrations by part:
\begin{eqnarray}
\sigma_1(t)&=&\frac{\mu_1\tau_1^3}{3}\int_{-\infty}^{t} e^{\left(-\tau_1^{1/3}(t-u)\right)}\epsilon (u)\, du\nonumber\\
&+&\frac{2\mu_1\tau_1^{1/3}}{3}\int_{t}^{+\infty} e^{\left(-\frac{\tau_1^{1/3}}{2}(u- t)\right)}\sin\left(\frac{\tau_1^{1/3}\sqrt{3}}{2}(u-t) + \frac{\pi}{6}\right)\epsilon (u)\, du\nonumber\\
&=& \sigma_{1*}(t)+\sigma_{1**}(t)
\end{eqnarray}
There are two contributions to the global stress $\sigma_1(t)$ due to the harmonic process. 
These contributions have two different relaxation times. The first one satisfies the equation
\begin{equation}
\frac{d\sigma_{1*}}{dt}(t)+\tau_1^{1/3}\sigma_*(t)=\frac{1}{3}\mu_1\tau_1^{1/3}\epsilon(t)
\end{equation}
characteristic of a Kelvin-like Model, while the second follows a higher order differential equation with stress and strain derivatives:
\begin{equation}
\frac{d^3\sigma_{1**}}{dt^3}(t)+\tau_1\sigma_{1**}(t)=\frac{2}{3}\mu_1\tau_1\epsilon(t)+\frac{1}{3}\mu_1\tau_1^{2/3}\frac{d\epsilon}{dt}(t)-\frac{1}{3}\mu_1\tau_1^{1/3}\frac{d^2\epsilon}{dt^2}(t)
\end{equation}
}
\AT{Let's now consider the {\bf anharmonic contribution} $G_2(\omega)$. The inverse Fourier Transforms gives the real-time stress that can be integrated by part:
\begin{eqnarray}
\sigma_2(t)&=&\frac{1}{2\pi}\int_{-\infty}^{+\infty}\left\lbrack\frac{\mu_2\tau_2^2}{\omega^2+\tau_2^2}\hat{\epsilon}(\omega)+\frac{\tau_2\mu_2}{\omega^2+\tau_2^2}i\omega\hat{\epsilon}(\omega)\right\rbrack\exp{(i\omega t)}\,d\omega\nonumber\\
&=&\frac{\mu_2\tau_2}{2}\int_{-\infty}^{+\infty}e^{-\tau_2\vert t-u\vert}\epsilon(u)\,du+\frac{\mu_2}{2}\int_{-\infty}^{+\infty}e^{-\tau_2\vert t-u\vert}\frac{d\epsilon}{du}(u)\,du\nonumber\\
&=&\mu_2\tau_2\int_t^{+\infty}\epsilon(u)e^{-\tau_2(u-t)}\,du
\end{eqnarray}
giving rise to the following Kelvin-like equation:
\begin{equation}
\epsilon(t)=\frac{1}{\mu_2}\sigma_2(t)-\frac{1}{\mu_2\tau_2}\frac{d\sigma_2}{dt}(t)
\end{equation}
In comparison to the initial Maxwell Model, the viscous stress is replaced in this model by a second order temporal stress derivative. 
}

\AT{\section*{Appendix C: Generalization to other frequency dependence of viscosities}}

\AT{Let's consider now the general power-law case. In this case, the frequency dependent viscosities are written more generally $\eta_1^*=\eta_1\omega^{\alpha_1}$ and $\eta_2^*=\eta_2\omega^{\alpha_2}$. The internal friction then becomes:
\begin{eqnarray}
\frac{G"}{G'}&=&\frac{G_1"+G_2"}{G_1'+G_2'}=\frac{\frac{\mu_1^2\eta_1\omega^{\alpha_1+1}}{\mu_1^2+\eta_1^2\omega^{2\alpha_1+2}}+\frac{\mu_2^2\eta_2\omega^{\alpha_2+1}}{\mu_2^2+\eta_2^2\omega^{2\alpha_2+2}}}{\frac{\mu_1\eta_1^2\omega^{2\alpha_1+2}}{\mu_1^2+\eta_1^2\omega^{2\alpha_1+2}}+\frac{\mu_2\eta_2\omega^{2\alpha_2+2}}{\mu_2^2+\eta_2^2\omega^{2\alpha_2+2}}}\label{eq:frictionAppC}\\
&=&\frac{\mu_1^2\eta_1\mu_2^2\omega^{\alpha_1+1}+\mu_1^2\eta_1\eta_2^2\omega^{\alpha_1+2\alpha_2+3}+\mu_1^2\mu_2^2\eta_2\omega^{\alpha_2+1}+\mu_2^2\eta_2\eta_1^2\omega^{\alpha_2+2\alpha_1+3}}{\mu_1\eta_1^2\mu_2^2\omega^{2\alpha_1+2}+\mu_1\eta_1^2\eta_2^2\omega^{2\alpha_1+2\alpha_2+4}+\mu_2\eta_2^2\mu_1^2\omega^{2\alpha_2+2}+\mu_2\eta_1^2\eta_2^2\omega^{2\alpha_1+2\alpha_2+4}}\nonumber
\end{eqnarray}
In the following, we will assume that {\bf $\alpha_1<\alpha_2\leq -1$}.}

\AT{{\bf C.1. Asymptotic behaviour $\omega\rightarrow 0$}:}

\AT{The low frequency limit of the numerator is $\propto \mu_2^2\eta_2\eta_1^2\omega^{\alpha_2+2\alpha_1+3}$, while the denominator in this limit is $\propto (\mu_1+\mu_2)\eta_1^2\eta_2^2\omega^{2\alpha_2+2\alpha_1+4}$. They both result from the product of the two viscous terms in Eq.~(\ref{eq:frictionAppC}). The low frequency limit of the internal friction is thus
\begin{equation}
\lim_{\omega\rightarrow 0}\frac{G"}{G'}\approx\frac{\mu_2}{(\mu_1+\mu_2)\eta_2}\omega^{-\alpha_2-1}
\end{equation}
The power-law in the low frequency limit thus depends only on $\alpha_2$, with a prefactor involving the rigidity $\mu_1$ of the parallel setup.}

\AT{{\bf C.2. Asymptotic behaviour $\omega\rightarrow\infty$}:}

\AT{The high frequency limit of the numerator is $\propto \mu_1^2\mu_2^2\eta_2\omega^{\alpha_2+1}$, while the denominator in this limit is $\propto\mu_2\eta_2^2\mu_1^2\omega^{2\alpha_2+2}$. This results from the product of the dominant rigid elastic part of the setup 1 with the viscous part of the setup 2 in Eq.~\ref{eq:frictionAppC}. The high frequency limit of the internal friction becomes
\begin{equation}
\lim_{\omega\rightarrow +\infty}\frac{G"}{G'}\approx\frac{\mu_2}{\eta_2}\omega^{-\alpha_2-1}
\end{equation}
The power-law frequency dependence of the internal friction in the high frequency limit thus depends only on the characteristics of the setup with the power-law with the smallest absolute value ($\alpha_2$, with $\alpha_2=-2$ in our model).
}

\AT{{\bf C.3. Intermediate frequency range}: }

\AT{In this frequency range, the internal friction becomes
\begin{equation}
\frac{G"}{G'}\approx\frac{\mu_1^2\eta_1\mu_2^2\omega^{\alpha_1+1}+\mu_1^2\eta_1\eta_2^2\omega^{\alpha_1+2\alpha_2+3}}{\mu_1\eta_1^2\mu_2^2\omega^{2\alpha_1+2}}\propto\omega^{2\alpha_2-\alpha_1+1}\textrm{ and then }\propto\omega^{-\alpha_1-1}
\end{equation}
The intermediate regime is thus controlled by the power-law with the largest absolute value ($\alpha_1$, with $\alpha_1=-4$ in our model).}

\section*{Acknowledgement}
This work benefited form the early participation of Robin Rouzaud. Interesting discussions with Giulio Monaco are acknowledged. H.L. is financed by the french ministry of research. A.T., H.L. and A.G. thank ECOS chile - grant No. C17E02 for their support.


\end{document}